\documentclass[aps,pra,twocolumn,notitlepage,superscriptaddress,showpacs]{revtex4-1}

\usepackage{amsmath}
\usepackage{amsfonts}
\usepackage{amsthm}
\usepackage[utf8]{inputenc}
\usepackage{graphicx}
\usepackage{color}
\usepackage{quantum}
\usepackage[colorlinks]{hyperref}
	\hypersetup{
		bookmarksnumbered,
		pdfstartview={FitH},
		citecolor={blue},
		linkcolor={red},
		urlcolor={black},
		pdfpagemode={UseOutlines}
	}

\newcommand{\bg}[1]{\boldsymbol{#1}}
\newcommand{\inlineheading}[1]{\textit{{#1.---}}}
\newcommand{\mc}[1]{\mathcal{#1}}
\newcommand{\mb}[1]{\mathbf{#1}}
\newcommand{\nf}{N_\mc{F}}
\newcommand{\nlj}{N_{LJ}}
\newcommand{\nljt}{\tilde{N}_{LJ}}
\newcommand{\trnorm}[1]{\| #1 \|_{\tr}}

\newtheorem{thm}{Theorem}

\newtheorem{obs}{Observation}

\begin{document}
\title{A general framework for quantum macroscopicity in terms of coherence}

\author{Benjamin Yadin} \affiliation{Atomic and Laser Physics, Clarendon Laboratory, University of Oxford, Parks Road, Oxford, OX1 3PU, UK}
\author{Vlatko Vedral} \affiliation{Atomic and Laser Physics, Clarendon Laboratory, University of Oxford, Parks Road, Oxford, OX1 3PU, UK} \affiliation{Centre for Quantum Technologies, National University of Singapore, Singapore 117543}
\date{\today}
\begin{abstract}
We propose a universal language to assess macroscopic quantumness in terms of coherence, with a set of conditions that should be satisfied by any measure of macroscopic coherence. We link the framework to the resource theory of asymmetry. We show that the quantum Fisher information gives a good measure of macroscopic coherence, enabling a rigorous justification of a previously proposed measure of macroscopicity. This picture lets us draw connections between different measures of macroscopicity and evaluate them; we show that another widely studied measure fails one of our criteria.
\end{abstract}


\maketitle

\inlineheading{Introduction}
One of the most difficult challenges in quantum theory is to explain how it can be compatible with the classical world observed at the macroscopic scale. We are accustomed to the fact that microscopic systems can exist in superposition states, but quantum theory also allows this behaviour in principle at the macroscopic scale. Various attempts have been made to address this; one common approach may be broadly termed `decoherence', in which the quantumness of a large system is highly susceptible to being destroyed by interactions with a noisy environment \cite{zurek2003decoherence}; similarly, it has been argued that classical behaviour emerges from limitations on the precision of measurements \cite{kofler2007classical,sekatski2014difficult,raeisi2011coarse}. Others have suggested fundamental modifications to quantum theory that are negligible at microscopic scales but cause macroscopic superpositions to quickly appear classical \cite{bassi2013models}.

Ultimately, it is up to experiments to probe the boundary between the quantum and classical worlds, and much progress has been made in this direction in recent decades \cite{leggett2002testing,arndt2014testing,farrow2015classification}. A wide variety of systems have been explored, including interference of large molecules \cite{nairz2003quantum,eibenberger2013matter}, superpositions of coherent states in photonic systems \cite{gao2010experimental}, superconducting circuits behaving as large-scale qubits \cite{vanderwal2000quantum}, and the control of low-lying vibrational states of micromechanical oscillators \cite{oconnell2010quantum}.

Given this diversity, we need a general way to quantify how well each experiment has achieved its aim of creating large-scale quantum coherence. Beginning with ideas by Leggett \cite{leggett1980macroscopic,leggett2002testing}, a variety of measures of `quantum macroscopicity' have been proposed, each motivated differently and aiming to capture a different potentially macroscopic quantum property. Some assume that the given state is a superposition of the form $\ket{\psi_0}+\ket{\psi_1}$ and quantify to what extent the two branches are macroscopically distinct \cite{marquardt2008measuring,korsbakken2007measurement,sekatski2014size} or useful for interferometry \cite{bjork2004size}. Others apply to more general states of systems of many qubits \cite{frowis2012measures}, continuous-variable modes \cite{lee2011quantification} and mechanical objects \cite{nimmrichter2013macroscopicity}.

These various approaches do not share any precise unifying principles. Inspired by recent work on quantifying coherence \cite{baumgratz2014quantifying,aberg2006quantifying}, we propose a mathematical framework of very general applicability for defining macroscopic coherence and give criteria to be satisfied by measures. This aims to provide a rigorous meaning to the notion of a macroscopic superposition. We show that this is mathematically equivalent to an instance of the resource theory of asymmetry, which describes the available states and operations under symmetry constraints. Via this approach, we give a solid motivation to a previously proposed measure of macroscopicity based on the quantum Fisher information, and show that a particular measure based on phase space structure fails one of the criteria.

\inlineheading{Defining the problem}
Let us suppose that, for a given experiment of the kind mentioned above, there is some particular `macroscopic observable' $A$ of interest. This is always a quantity which makes sense to use for the classical macroscopic scale description of the system: the centre of mass for molecular interference or micromechanical oscillators; a field quadrature for a photonic cat state; the current for a superconducting circuit. We can say that such an experiment has created macroscopic coherence when there is a superposition of states with macroscopically different values of $A$. Thus we must first be able to quantify coherence associated with a superposition of states differing from one another in $A$-value by a fixed amount $\delta$; we will later call this $\delta$-coherence.

There has recently been progress in rigorously quantifying coherence (with respect to some fixed basis) using the framework of resource theories \cite{baumgratz2014quantifying}. In general resource theories, `free states' are states containing no resource and `free operations' are those which are unable to create the resource. When the resource is coherence, the free (i.e., incoherent) states are all diagonal states in the preferred basis; the free (i.e., incoherent) operations are those which take any incoherent state to another incoherent state. A measure of a general resource must vanish for free states and never increase under free operations -- thus measures of coherence can be defined.

One might attempt to define macroscopic coherence in this way, with reference to the eigenbasis $\ket{i}$ of the observable $A = \sum_i a_i \ketbra{i}{i}$. However, this formalism is not appropriate: the incoherent operations permit arbitrary permutations of the basis states, allowing for instance a `microscopic superposition' $(\ket{i}+\ket{j})/\sqrt{2}$ with a small difference $\delta=a_i-a_j$ to be turned into one with a macroscopic value of $\delta$. In short, the resource theory of coherence is ignorant of the values $a_i$ of the observable.

To rectify this, we introduce the notion of $\delta$-coherence to be coherence associated with a superposition of $\ket{i}$ and $\ket{j}$ such that $a_i-a_j = \delta$. Formally, by expanding a state $\rho = \sum_{i,j} \rho_{i,j} \ketbra{i}{j}$, we can identify the part containing all the $\delta$-coherence as
\begin{equation} \label{eqn:mode_decomp}
	\rho^{(\delta)} := \sum_{i,j:\, a_i-a_j=\delta} \rho_{i,j} \ketbra{i}{j}.
\end{equation}
The entire state can be expressed as $\rho = \sum_{\delta \in \Delta} \rho^{(\delta)}$, where $\Delta := \{a_i - a_j\}_{i,j}$ is the set of all eigenvalue differences of $A$.

We consider the amounts of $\delta$-coherence for each $\delta$ to all be resources of independent interest. Hence we define the free operations to be such that $\delta$-coherence cannot be created from a state with none initially. To do this, we must impose that $\delta$-coherence cannot be turned into $\delta'$-coherence for $\delta \neq \delta'$ -- without this constraint, a state with only $\delta$-coherence could be turned into one with nonzero $\delta'$-coherence. Thus a free operation $\mc{E}$ is any quantum operation (it need not be trace-preserving, as we allow nondeterministic operations) such that
\begin{equation} \label{eqn:free_ops}
	\mc{E}(\rho)^{(\delta)} = \mc{E}(\rho^{(\delta)}) \; \forall \delta \in \Delta.
\end{equation}

As we have seen, in accordance with resource theories, we demand several properties of a measure $\mc{M}$. Firstly, it must be nonnegative and vanish for states with no coherence:
\begin{description}
	\item[(M1)] $\mc{M}(\rho) \geq 0$ and $\mc{M}(\rho) = 0 \Leftrightarrow \rho = \rho^{(0)}$.
\end{description}
Furthermore, $\mc{M}$ should not increase under a free operation -- and if there are various possible outcomes, it should not increase on average:
\begin{description}
	\item[(M2a)] For a deterministic free operation $\mc{E}$ (such that $\tr \mc{E}(\rho) = 1$), $\mc{M}(\rho) \geq \mc{M}(\mc{E}(\rho))$
	\item[(M2b)] For $\mc{E} = \sum_\alpha \mc{E}_\alpha$ such that each $\mc{E}_\alpha$ satisfies (\ref{eqn:free_ops}), $\mc{M}(\rho) \geq \sum_\alpha p_\alpha \mc{M}(\sigma_\alpha)$, where the outcome $\sigma_\alpha = \mc{E}_\alpha(\rho)/p_\alpha$ occurs with probability $p_\alpha = \tr \mc{E}_\alpha(\rho)$.
\end{description}
Measures satisfying these conditions are said to be monotones. Another condition which one often demands is convexity, meaning that a measure does not increase under mixing:
\begin{description}
	\item[(M3)] $\mc{M}(\sum_i p_i \rho_i) \leq \sum_i p_i \mc{M}(\rho_i)$.
\end{description}
These conditions should be familiar from properties of entanglement measures, where the free states are separable and the free operations are LOCC (local operations and classical communication). They are not all independent -- (M2b) and (M3) together imply (M2a).

The theory we have built up is actually a special case of the resource theory of quantum reference frames, also known as asymmetry \cite{bartlett2007reference}. Our free operations are symmetric (or covariant) with respect to the transformations $T_x(\rho) := e^{-ixA} \rho e^{ixA},\, x \in \mathbb{R}$, meaning that $\mc{E}(T_x(\rho)) = T_x(\mc{E}(\rho))$.

Equation (\ref{eqn:mode_decomp}) is called a decomposition into modes of asymmetry in \cite{marvian2014modes}. As noted there, the trace norm $\trnorm{X} := \tr(\sqrt{X^\dagger X})$ provides a set of monotones $\trnorm{\rho^{(\delta)}}$ satisfying (M2). This in fact works for any norm $\| \cdot \|$ which is contractive under trace-preserving operations, meaning $\|\mc{E}(X)\| \leq \|X\|$, and convex. Then $\|\rho^{(\delta)}\|$ is a measure of the $\delta$-coherence in $\rho$. (A weaker form of (M1) is satisfied, as this vanishes when the state has no $\delta$-coherence.)

The concept of $\delta$-coherence answers the question `how much coherence does a state have at a scale $\delta$?' To quantify macroscopic coherence, we would like a measure taking all scales into account, giving higher weight to larger $\delta$. For this we suggest an additional condition imposing a strict order on the simplest superpositions:
\begin{description}
	\item[(M4)] Let $\ket{\psi} = \frac{1}{\sqrt{2}} (\ket{i} + \ket{j})$ and $\ket{\phi} = \frac{1}{\sqrt{2}} (\ket{k} + \ket{l})$. If $\abs{a_i - a_j} > \abs{a_k - a_l}$ then $\mc{M}(\ketbra{\psi}{\psi}) > \mc{M}(\ketbra{\phi}{\phi})$.
\end{description}

\inlineheading{Constructing measures}
We now use some results derived in asymmetry to construct measures satisfying (M1-4). The first result applies to the variance $V(\ket{\psi},A) := \braXket{\psi}{A^2}{\psi} - \braXket{\psi}{A}{\psi}^2$:
\begin{thm} \label{lem:variance_monotone}
	$V(\ket{\psi},A)$ satisfies (M1), (M2a), (M2b) and (M4) for pure states.
	\begin{proof}
		Lemma 8 in \cite{gour2008resource} proves (M1), (M2a) and (M2b) for the case of $U(1)$ reference frames. Simple modifications suffice to prove the slightly more general case considered here (see Appendix \ref{app:variance}). (M4) is clear from $V((\ket{i}+\ket{j})/\sqrt{2},A) = \frac{1}{4}(a_i - a_j)^2$.
	\end{proof}
\end{thm}

An extension to mixed states is achieved by the quantum Fisher information \cite{braunstein1994statistical}. In its most general form the quantum Fisher information $\mc{F}(\rho_x)$ is defined for a family of states $\rho_x$ depending on a continuous real parameter $x$ (or possibly multiple parameters), and evaluated at some $x_0$. Then $\left. \mc{F}(\rho_x) \right|_{x=x_0} := 4 \left. \partial_x^2 D_B^2(\rho_{x_0},\rho_x) \right|_{x=x_0}$, in terms of the Bures distance $D_B(\rho,\sigma) := 2(1 - F(\rho,\sigma))$, where $F(\rho,\sigma) = \tr \sqrt{\sqrt{\rho}\sigma\sqrt{\rho}}$ is the fidelity. Thus $\mc{F}(\rho_x)$ measures the rate of change of the state with respect to $x$.

We need the Fisher information measuring the sensitivity of the state under the transformations $T_x$, $\mc{F}(\rho, A) := \left. \mc{F}( e^{-ixA} \rho e^{ixA}) \right|_{x=0}$. This can be expressed as
\begin{equation} \label{eqn:fisher_expansion}
	\mc{F}(\rho, A) = 2 \sum_{a,b} \frac{(\lambda_a - \lambda_b)^2}{\lambda_a + \lambda_b} \abs{\braXket{\psi_a}{A}{\psi_b}}^2,
\end{equation}
where $\rho = \sum_a \lambda_a \ketbra{\psi_a}{\psi_a}$ is a spectral decomposition and the sum is over all $a,\,b$ such that $\lambda_a + \lambda_b >0$. For pure states, $\mc{F}(\ketbra{\psi}{\psi},A) = 4V(\ket{\psi},A)$.

\begin{thm} \label{thm:fisher_monotone}
	The quantum Fisher information $\mc{F}(\rho, A)$ satisfies (M1-4) for all states.
	\begin{proof}
		Any state $\rho$ can be decomposed in many ways as a mixture of pure states: $\rho = \sum_\mu p_\mu \ketbra{\psi_\mu}{\psi_\mu}$, where the $p_\mu$ form a probability distribution (the $\ket{\psi_\mu}$ need not be orthogonal). This lets us do a convex roof construction to extend any real-valued function $f$ of pure states to mixed states: $f_{CR}(\rho) := \inf_{ \{ p_\mu, \ket{\psi_\mu}  \} }\, \sum_\mu p_\mu f(\ket{\psi_\mu})$, where the optimisation is over all possible such decompositions. By construction, this is convex and reduces to $f$ for pure states.
		
		It was shown in \cite{toloui2011constructing} in the context of asymmetry that the convex roof of any pure state measure satisfying (M1-2) is a monotone on all states. Given Lemma \ref{lem:variance_monotone}, the convex roof of the variance is therefore a monotone. This quantity was called the `frameness of formation' in \cite{toloui2011constructing}, but it has actually been shown more recently that $\frac{1}{4}\mc{F}(\rho,A)$ is the convex roof of $V(\ket{\psi},A)$ \cite{toth2013extremal,yu2013quantum}.
	\end{proof}
\end{thm}

Note that other asymmetry monotones satisfying (M1-3) exist, such as the relative entropy measure $\min_{\sigma \in \mc{I}}S(\rho || \sigma)=\min_{\sigma \in \mc{I}} \tr(\rho \log \rho - \rho \log \sigma)$, where $\mc{I}$ is the set of incoherent states. However, this does not depend on the eigenvalues of $A$, so fails (M4). The Wigner-Yanase-Dyson skew information \cite{wigner1963information} $I_{sk}(\rho,A):=-\frac{1}{2}\tr([\sqrt{\rho},A]^2)$ is also a monotone \cite{marvian2014extending,girolami2014observable} and satisfies (M4), but is essentially equivalent to $\mc{F}$ in that $I_{sk}(\rho,A) \leq \mc{F}(\rho,A) \leq 2I_{sk}(\rho,A)$ \cite{petz2011introduction}.

The theorem below, following \cite{gour2008resource,toloui2011constructing}, demonstrates the meaning of $\mc{F}$ in terms of the equivalence of many copies of a given state with many copies of a chosen reference state.

\begin{thm} \label{thm:asymptotic}
	Let $\ket{\phi}$ be a reference state with $V(\ket{\phi},A)=A_0$. Coherence for $n$ copies is defined via the observable $\sum_{i=1}^n A_i$. Then in the limit of large $n$, $\ket{\psi}^{\ox n}$ and $\ket{\phi}^{\ox m}$ have the same $\delta$-coherence for all $\delta$, such that $m/n = V(\ket{\psi},A)/A_0$.
	
	For mixed states, the minimal average ratio $m/n$ over all pure state decompositions $\rho = \sum_\mu p_\mu \ketbra{\psi_\mu}{\psi_\mu}$ is $\mc{F}(\rho,A)/(4A_0)$.
	
	\begin{proof}
		See Appendix \ref{app:asymptotic} for details. The idea is that the central limit theorem guarantees that the distribution of $\sum_{i=1}^n A_i$ tends to a normal distribution which is characterised fully by its mean and variance. The $\delta$-coherence depends only on this distribution; it is independent of the mean but proportional to the number of copies. The final statement follows from the convex roof property.
	\end{proof}
\end{thm}

\inlineheading{Evaluating existing measures}
\emph{i) The measure by Fr\"owis and D\"ur}: In \cite{frowis2012measures}, the authors compared various measures of macroscopicity for systems of $N$ qubits \cite{bjork2004size,korsbakken2007measurement,marquardt2008measuring,shimizu2005detection}. They proposed a measure based on the quantum Fisher information:
\begin{equation} \label{eqn:nf}
	\nf(\rho) := \max_{A \in \mc{A}}\, \frac{1}{4N} \mc{F}(\rho, A),
\end{equation}
where $\mc{A}$ is the set of all observables that can be written as $A = \sum_{i=1}^N A_i$ over local $A_i$, each acting on a single qubit $i$ and with bounded spectrum ($\| A_i \| = 1$ for some norm) \footnote{$\mc{A}$ may be extended to include `$k$-local' $A_i$ acting on groups of $k$ qubits, with $k$ bounded independent of $N$, which case the denominator of (\ref{eqn:mode_decomp}) contains the number of groups in place of $N$. This modification lets the measure capture correlations between groups of sites.}. Thought of as a total spin, these could reasonably be described as macroscopic observables.

This measure was motivated by its ability to capture some supposedly macroscopic quantum properties. For instance, $\nf > k-1$ indicates genuine $k$-body entanglement \cite{toth2012multipartite,hyllus2012fisher}. Fisher information is also a central quantity in quantum metrology \cite{paris2009quantum}. $\mc{F}(\rho,A) = O(N^2)$ enables an unknown parameter $x$ encoded in a state via $T_x$ to be estimated with precision $\Delta x \propto 1/N$ compared with the classical `shot-noise' scaling $\Delta x \propto 1/\sqrt{N}$ for $\mc{F}(\rho,A) = O(N)$. So $\nf$ measures the highest quantum improvement of a given state over the classical bound for estimating transformations $T_x$ generated by any observable in $\mc{A}$.

$\nf$ has been called an `effective size'; it is supposed to measure the size up to which quantum properties are significant. Our framework helps justify $\nf$ more rigorously as a measure of macroscopicity:

\begin{obs}
	The quantity $\nf$ measures the maximum macroscopic coherence, according to the quantum Fisher information, over all observables in $\mc{A}$.
\end{obs}

Note that if we take the reference state in Theorem \ref{thm:asymptotic} to be a product of single-qubit states $\bigotimes_{i=1}^N \ket{\phi_i}$ with $V(\ket{\phi_i},A_i)=1$, then the average ratio of copies is exactly $\mc{F}(\rho,A)/(4N)$.

This measure could in principle be applied to any physical system, as long as one can convincingly choose a set $\mc{A}$ of macroscopic observables. To illustrate this, we shall look at an example. For $N$ continuous-variable bosonic modes with annihilation and creation operators $a_i$ and $a^\dagger_i$, it has been suggested to maximize the Fisher information over sums of quadratures \cite{frowis2014linking,oudot2014two}. Then the set $\mc{A}$ contains  $x^{\bg{\theta}} := \sum_{i=1}^N x_i^{\theta_i}$, where $\bg{\theta}:=\{\theta_1,\dots,\theta_N\}$ and $x_i^{\theta_i} := \cos(\theta_i)x_i + \sin(\theta_i)p_i$ is a rotated quadrature of the $i$th mode with associated conjugate quadratures satisfying the canonical commutation relation $[x_i,p_i]=i$.

For a single mode and a fixed quadrature $x$, $V(\ket{\psi},x)$ is the (squared) dispersion in $x$ of the phase space distribution of $\ket{\psi}$. The free operations also have a simple interpretation using Lemma 2 of \cite{gour2008resource}: they can be constructed from phase space displacements, $\rho \to e^{\alpha a^\dagger - \alpha^* a} \rho e^{-\alpha a^\dagger + \alpha^* a}$, and generalized measurements of $x$, $\rho \to M \rho M^\dagger$ with $M = \int dx \, f(x) \ketbra{x}{x}$. A reasonable choice of reference state for Theorem \ref{thm:asymptotic} is any coherent state $\ket{\alpha}$ (defined by $a \ket{\alpha} = \alpha \ket{\alpha}$), which has many classical properties. Since $V(\ket{\alpha},x)=1/2$, we find that $n$ copies of $\ket{\psi}$ are equivalent to $n V(\ket{\psi},x)/2$ copies of $\ket{\alpha}$.

Maximizing over quadratures, $\nf(\ketbra{\psi}{\psi})$ is the largest dispersion over all directions in phase space. For a number eigenstate $\ket{n}$, a superposition of coherent states $\ket{\alpha}+\ket{-\alpha}$ and a quadrature-squeezed coherent state $e^{(\xi^* a^2 - \xi {a^\dagger}^2)/2}\ket{\alpha}$ -- all typically described as nonclassical states -- it is simple to show that $\nf$ scales approximately with the expected number of particles.

\emph{ii) The measure by Lee and Jeong} \cite{lee2011quantification}: This measure for systems of $N$ modes aims to quantify nonclassicality associated with oscillations in the Wigner function in phase space \cite{gerry2005introductory}. We define a displacement operator $D(\bg{\alpha}) := \prod_i e^{\alpha_i a_i^\dagger - \alpha_i^* a_i}$, where $\bg{\alpha} = (\alpha_1,\alpha_2,\dots,\alpha_N)$. We write their measure as
\begin{equation} \label{eqn:I_LJ}
	\nlj(\rho) := \frac{1}{2\pi^N} \int d^{2N} \bg{\alpha} \; \abs{\bg{\alpha}}^2 \abs{\chi_\rho(\bg{\alpha})}^2,
\end{equation}
where $\abs{\bg{\alpha}}^2 := \sum_{i=1}^N \abs{\alpha_i}^2$ and $\chi_\rho(\bg{\alpha}) := \tr[\rho D(\bg{\alpha})]$ is the characteristic function \footnote{Note that their original definition instead used $|\bg{\alpha}|^2 - N$, but removing the constant shift ensures that the measure is nonnegative \cite{jeong2011reply}}.

\begin{thm} \label{thm:I_LJ}
	The measure $\nlj$ has the following properties:
	\begin{enumerate}
		\item[(a)] It measures the rate of decrease of purity under a certain decoherence model: $\nlj(\rho) = -\frac{1}{2} \partial_t \tr(\rho^2)$ with $\partial_t \rho = \mc{L}(\rho) := -\frac{1}{4} \sum_{i=1}^N [x_i,[x_i,\rho]] + [p_i,[p_i,\rho]]$.
		\item[(b)] Define the quantity $I_L(\rho,A) := -\frac{1}{2} \tr([\rho,A]^2)$, which was proposed as an observable lower bound to coherence in \cite{girolami2014observable}. Then $\nlj(\rho) = \frac{1}{2} \sum_{i=i}^N I_L(\rho, x_i) + I_L(\rho, p_i)$.
		\item[(c)] $\frac{1}{N} \nlj(\rho) \leq \nf(\rho) := \max_{\bg{\theta}} \, \frac{1}{4N} \mc{F}(\rho, x^{\bg{\theta}})$.
	\end{enumerate}
	\begin{proof}
		Part (a) follows from a direct calculation that shows $\partial_t \chi(\bg{\alpha}) = -\frac{1}{2} \abs{\bg{\alpha}}^2 \chi(\bg{\alpha})$ (see Appendix \ref{app:decoherence}), while (b) is a simple matter of evaluating $\nlj(\rho) = -\tr[\rho \mc{L}(\rho)]$ and using $\tr(\rho [A,[A,\rho]]) = \tr([\rho, A]^2)$. For (c), it suffices to write $I_L(\rho,A) = \frac{1}{2} \sum_{a,b} (\lambda_a - \lambda_b)^2 \abs{\braXket{\psi_a}{A}{\psi_b}}^2$ and compare with (\ref{eqn:fisher_expansion}) to see that $I_L(\rho,A)\leq \frac{1}{4}\mc{F}(\rho,A)$. It is easy to check that $\mc{L}$ is isotropic in the sense that $x_i,\,p_i$ can be replaced with $x_i^{\theta_i},\,x_i^{\theta_i+\pi/2}$ for any $\theta_i$. Then $\nlj(\rho) \leq \max_{\bg{\theta}}\, I_L(\rho,x^{\bg{\theta}}) \leq \max_{\bg{\theta}}\, \frac{1}{4}\mc{F}(\rho, x^{\bg{\theta}})$.
	\end{proof}
\end{thm}
The original paper related the measure to a decoherence model slightly different from that in (a). We note in passing that this provides a connection with a measure based on the rate of decoherence under an intrinsic collapse model in mechanical phase space \cite{nimmrichter2013macroscopicity}. As shown in Appendix \ref{app:nh_model}, in a natural limit their decoherence model for a single massive object is $\partial_t \rho \approx - c_x [x,[x,\rho]] - c_p [p,[p,\rho]]$, where $c_x,\, c_p$ are coefficients. However, the measure in \cite{nimmrichter2013macroscopicity} is not a straightforward function of the state, so we cannot make precise statements about (M1-4).

As in \cite{frowis2014linking}, we see from (b) that $\nlj$ is additive with respect to modes and hence cannot capture correlations -- this explains the $1/N$ normalisation in (c). To remedy this problem, while putting the measure on an equal footing with $\nf$ by using a maximization over observables, we define a refined version by
\begin{equation} \label{eqn:IPrime_LJ}
	\nljt(\rho) := \max_{\bg{\theta}}\, \frac{1}{N} I_L(\rho, x^{\bg{\theta}}),
\end{equation}
satisfying $\nljt(\rho) \leq \nf(\rho)$. Moreover, $\nljt(\ketbra{\psi}{\psi}) = \nf(\ketbra{\psi}{\psi})$ for any pure state $\ket{\psi}$ because $I_L$, like $\mc{F}/4$, then equals $V$.

To determine whether $\nljt$ is a good measure of macroscopicity in the same way as $\nf$, we must ask if $I_L$ satisfies (M1-4). The following answers this in the negative:
\begin{obs}
	$I_L$ can increase under the free operations defined by (\ref{eqn:free_ops}).
	\begin{proof}
		Suppose we have the system and an ancilla in a state $\rho \ox \sigma$ with $\tr(\sigma^2)<1$. Then $I_L(\rho \ox \sigma, A \ox I) = I_L(\rho,A) \tr(\sigma^2)$. Thus the measure increases by replacing the ancilla with some pure state, which is a free operation with respect to $A \ox I$.
	\end{proof}
\end{obs}
This problem stems from the relation of $I_L$ to the Hilbert-Schmidt distance $D_{HS}(\rho, \sigma) := \tr[(\rho-\sigma)^2]$. We have $I_L(\rho,A) = \frac{1}{2} \left. \partial_x^2 D_{HS}(\rho, e^{-ixA} \rho e^{ixA}) \right|_{x=0}$, as for $\mc{F}$ and $D_B$. The distance $D_{HS}$ is problematic when used to define various quantum information-theoretic quantities, notably measures of entanglement and discord, because it is not contractive under trace-preserving operations \cite{ozawa2000entanglement,piani2012problem}. This cannot be fixed by simply dividing by the purity, as it leads to a poor characterisation of mixed state superpositions \cite{jeong2011reply}.

We finally address whether $I_L$ is a faithful lower bound to $\mc{F}$. This is true in the sense that $I_L(\rho,A)=0 \Leftrightarrow \mc{F}(\rho,A)=0$. However, for macroscopicity we are concerned with the scaling of the measure. We provide an $N$-qubit example where the two measures exhibit entirely different scaling.

\begin{obs}
	It is possible to have $I_L = O(N)$ but $\mc{F} = O(N^2)$.
	\begin{proof}
	Let $N$ be even and define $\rho_N := \sum_{k=0}^{N/2-1} \frac{2}{N} \ketbra{\psi^N_k}{\psi^N_k}$, where $\ket{\psi^N_k} := \frac{1}{\sqrt{2}} \left( \ket{0^{N-k}1^k} + \ket{1^{N-k}0^k} \right)$. Then $I_L(\rho_N, \sum_i \sigma^z_i) = O(N)$, $\mc{F}(\rho_N, \sum_i \sigma^z_i) = O(N^2)$. The calculations proceed from the expansion (\ref{eqn:fisher_expansion}) and the equivalent for $I_L$. See Appendix \ref{app_scaling} for details.
	\end{proof}
\end{obs}

\inlineheading{Conclusions}
In this Letter, we have introduced a formal framework for quantifying the performance of experiments that aim to create macroscopic quantum coherence. This may be seen as a more restricted version of the recent work on coherence monotones \cite{baumgratz2014quantifying}. We have demonstrated an equivalence with a type of asymmetry resource theory, enabling us to use existing results about monotones from that field. We have proved the quantum Fisher information to be a monotone and demonstrated its meaning in terms of the coherence of many copies of a state. This justifies an existing measure of quantum macroscopicity \cite{frowis2012measures} that uses the Fisher information, but was not previously motivated rigorously. We have compared this with another measure based on phase space structure \cite{lee2011quantification}, which can be seen as an attempt to quantify quadrature coherence. However, we have shown it is related to a measure which is not a monotone, and can significantly underestimate the Fisher information.

We expect our result to be applicable to a wide range of experiments that test the macroscopic limits of quantum mechanics, since it has a very general justification and the proposed measure is relatively simple to calculate. It should be possible to assess and compare a variety of existing systems, as in \cite{nimmrichter2013macroscopicity}, by taking comparable mechanical observables such as total momenta or currents.

There have been recent insights about ways to feasibly measure the quantum Fisher information \cite{girolami2014observable,frowis2015detecting,hauke2015measuring}, demonstrating that macroscopic coherence could be witnessed in the laboratory. \\

\vspace{0.5em}
The authors acknowledge funding from the National Research Foundation (Singapore), the Ministry of Education (Singapore), the EPSRC (UK), the Templeton Foundation, the Leverhulme Trust, the Oxford Martin School and Wolfson College, University of Oxford. The authors also thank Davide Girolami, Oscar Dahlsten, Raam Uzdin and Chiara Marletto for helpful discussions and comments.

\bibliographystyle{h-physrev}
\bibliography{macro_coherence}

\clearpage
\onecolumngrid
\appendix

\section{Characterisation of the free operations} \label{app:free_ops}
Let $\mc{E}$ be a completely positive trace-non-increasing operation with associated Kraus operators $K_\alpha$, such that $\mc{E}(\rho) = \sum_\alpha K_\alpha \rho K_\alpha^\dagger$ and $\sum_\alpha K_\alpha^\dagger K_\alpha \leq I$. Applying Lemma 1 of \cite{gour2008resource} to the case where $T_x$ forms a representation of either $U(1)$ or the real line under addition $\mathbb{R}_+$, we can always find a set of Kraus operators for $\mc{E}$ satisfying
\begin{equation} \label{eqn:kraus_condition}
T_x(K_\alpha) = e^{-ixA} K_\alpha e^{ixA} = e^{-ix\delta_\alpha} K_\alpha \quad \forall x \in \mathbb{R},
\end{equation}
for some set of $\delta_\alpha \in \mathbb{R}$. (The functions $e^{-ix\delta_\alpha}$ are irreducible representations of the relevant group.) Writing an arbitrary operator in the eigenbasis of $A$, we have
\begin{align}
K_\alpha & = \sum_{\delta \in \Delta} \sum_{(i,j)\sim\delta} c_{\alpha,i,j} \ketbra{i}{j}, \\
T_x(K_\alpha) & = \sum_{\delta \in \Delta} e^{-ix\delta} \sum_{(i,j)\sim\delta} c_{\alpha,i,j} \ketbra{i}{j},
\end{align}
where the notation $(i,j)\sim\delta$ means that $a_i-a_j=\delta$. Thus condition (\ref{eqn:kraus_condition}) can only be satisfied when $\delta_\alpha \in \Delta$ and then is equivalent to $K_\alpha = K_\alpha^{(\delta_\alpha)}$.

\section{Monotonicity of variance} \label{app:variance}
We will not repeat the proof of Lemma 8 in \cite{gour2008resource}, but instead note that the only point where a proof of our case differs slightly is in the evaluation of the commutator $[A,K_\alpha]$. The original assumes that $A = \hat{N}$ has a harmonic spectrum (its eigenvalues are all the nonnegative integers), and uses the fact that $[\hat{N},K_\alpha] = k_\alpha K_\alpha$ for some $k_\alpha \in \mathbb{Z}$.

In our case, as shown above, the Kraus operators can always be written in the form $K_\alpha = \sum_{(i,j)\sim\delta_\alpha} K_{\alpha,i,j} \ketbra{i}{j}$ for some $\delta_\alpha \in \Delta$. Then
\begin{align}
[A, K_\alpha] & = \sum_{(i,j)\sim\delta_\alpha} c_{\alpha,i,j} [A, \ketbra{i}{j}] \nonumber \\
& = \sum_{(i,j)\sim\delta_\alpha} c_{\alpha,i,j} (a_i-a_j) \ketbra{i}{j} \nonumber \\
& = \delta_\alpha K_\alpha,
\end{align}
and the remainder of the proof continues unaffected by the replacement $\hat{N} \to A$, $k_\alpha \to \delta_\alpha$.

\section{Asymptotic meaning of Fisher information} \label{app:asymptotic}
Our proof follows the ideas in \cite{schuch2004quantum,gour2008resource}. We take the observable of interest, $A$, to have a continuous spectrum (although the idea of the proof is the same in the discrete case). For any pure state we write $\ket{\psi} = \int \! dx \, \psi(x) \ket{\psi_x}$, such that each $\ket{\psi_x}$ is normalized by $\braket{\psi_x}{\psi_y} = \delta(x-y)$ and $A\ket{\psi_x} = x\ket{\psi_x}$. Then, for $n$ copies,
\begin{align}
\ket{\psi}^{\ox n} & = \int \! dX \, \sqrt{p_n(X)} \ket{\Psi_X} , \\
\sqrt{p_n(X)} \ket{\Psi_X} & := \int_{\sum_{x=1}^n x_i =X} \! d\mathbf{x} \, \bigotimes_{i=1}^n \psi(x_i) \ket{\psi_{x_i}} , \\
p_n(X) & = \int_{\sum_{x=1}^n x_i =X} \! d\mathbf{x} \, \prod_{i=1}^n \abs{\psi(x_i)}^2 ,
\end{align}
where $\mathbf{x} = (x_1,\dots,x_n)$. Now $p_n(X)$ is just the probability density that the values of the observables $A_i$ all sum to $X$, given that they individually have the distribution $\abs{\psi(x_i)}^2$. Under a set of reasonable assumptions, the central limit theorem says that as $n \to \infty$, $p_n(X)$ tends to a normal distribution with mean $n \braXket{\psi}{A}{\psi}$ and variance $\sigma^2 = n V(\ket{\psi},A)$. For the $\delta$-coherence, we have
\begin{equation}
\left( \ketbra{\psi}{\psi}^{\ox n} \right)^{(\delta)} = \int \! dX \, \sqrt{p_n(X+\delta) p_n(X)} \ketbra{\Psi_{X+\delta}}{\Psi_{X}}.
\end{equation}
If we now do the same for $m$ copies of the reference state $\ket{\phi}$, replacing the distribution $p_n(X)$ with $q_m(X)$, it is clear that for any norm, $\| ( \ketbra{\psi}{\psi}^{\ox n} )^{(\delta)} \| = \| ( \ketbra{\phi}{\phi}^{\ox m} )^{(\delta)} \|$ as long as $p_n(X) = q_m(X-X_0)$ with some constant shift $X_0$. This is guaranteed by the central limit theorem when $m/n = V(\ket{\psi},A) / V(\ket{\phi},A)$.

As in \cite{schuch2004quantum,gour2008resource}, we must note that if $\psi(x)$ or $\phi(x)$ vanish anywhere, then the application of the central limit theorem does not necessarily work. However, negligible amounts of extra resources suffice to remove these gaps.

\section{Decoherence model for $I_{LJ}$} \label{app:decoherence}
We first assemble some useful facts for a single mode \cite{gerry2005introductory}:
\begin{align}
\braket{\beta}{\alpha} & = e^{-\frac{1}{2}\abs{\beta-\alpha}^2 + \frac{1}{2}(\alpha \beta^* - \alpha^* \beta)}, \\
D(\alpha)D(\beta) & = e^{\frac{1}{2}(\alpha \beta^* - \alpha^* \beta)} D(\alpha + \beta), \\
\frac{1}{\pi} \int \! d^2\alpha \, \ketbra{\alpha}{\alpha} & = I.
\end{align}
Using these, we have
\begin{align}
\tr D(\alpha) & = \frac{1}{\pi} \int \! d^2\beta \, \braXket{\beta}{D(\alpha)}{\beta} \nonumber \\
& = \frac{1}{\pi} e^{-\abs{\alpha}^2} \int \! d^2\beta \, e^{\alpha \beta^* - \alpha^* \beta} \nonumber \\
& = \pi\, \delta^2(\alpha),
\end{align}
from which it follows that we can write (in the $N$-mode case)
\begin{align}
\rho = \frac{1}{\pi^N} \int \! d^{2N}\bg{\alpha} \; \chi(\bg{\alpha}) D(-\bg{\alpha}).
\end{align}
Under the model in Theorem 4a, using $[x_i,D(\bg{\alpha})] = \sqrt{2} \Re(\alpha_i) D(\bg{\alpha})$ and $[p_i,D(\bg{\alpha})] = \sqrt{2} \Im(\alpha_i) D(\bg{\alpha})$, we have
\begin{align}
\partial_t \rho & = -\frac{1}{4} \sum_{i=1}^N \frac{1}{\pi^N} \int \! d^{2N}\bg{\alpha} \; (2 \Re(\alpha_i)^2 + 2\Im(\alpha_i)^2) \chi(\bg{\alpha}) D(-\bg{\alpha}) \nonumber \\
& = -\frac{1}{2\pi^N} \int \! d^{2N}\bg{\alpha} \; \abs{\bg{\alpha}}^2 \chi(\bg{\alpha}) D(-\bg{\alpha}),
\end{align}
and so $\partial_t \chi(\bg{\alpha}) = -\frac{1}{2}\abs{\bg{\alpha}}^2 \chi(\bg{\alpha})$.

Finally, we calculate the rate of loss of purity:
\begin{align}
-\frac{1}{2} \partial_t \tr(\rho^2) & = - \tr(\rho\, \partial_t \rho) \nonumber \\
& = -\frac{1}{\pi^{2N}} \int \! d^{2N}\bg{\alpha}\, d^{2N}\bg{\beta}\; \chi(\bg{\alpha}) \partial_t \chi(\bg{\beta}) \, \tr[D(-\bg{\alpha})D(-\bg{\beta})] \nonumber \\
& = -\frac{1}{\pi^{2N}} \int \! d^{2N}\bg{\alpha}\, d^{2N}\bg{\beta}\; \chi(\bg{\alpha})^* \partial_t \chi(\bg{\beta}) \, \tr[D(\bg{\alpha})D(-\bg{\beta})] \nonumber \\
& = -\frac{1}{\pi^{2N}} \int \! d^{2N}\bg{\alpha}\, d^{2N}\bg{\beta}\; \chi(\bg{\alpha})^* \partial_t \chi(\bg{\beta}) \, e^{\frac{1}{2}(-\bg{\alpha}.\bg{\beta}^* + \bg{\alpha}^*.\bg{\beta})} \tr[D(\bg{\alpha}-\bg{\beta})] \nonumber \\
& = -\frac{1}{\pi^{2N}} \int \! d^{2N}\bg{\alpha}\, d^{2N}\bg{\beta}\; \chi(\bg{\alpha})^* \partial_t \chi(\bg{\beta}) \,e^{\frac{1}{2}(-\bg{\alpha}.\bg{\beta}^* + \bg{\alpha}^*.\bg{\beta})} \pi^N \, \delta^{2N}(\bg{\alpha}-\bg{\beta}) \nonumber \\
& = -\frac{1}{\pi^N} \int \! d^{2N}\bg{\alpha} \; \chi(\bg{\alpha})^* \partial_t \chi(\bg{\alpha}) \nonumber \\
& = \frac{1}{2\pi^N} \int \! d^{2N}\bg{\alpha} \; \abs{\bg{\alpha}}^2 \abs{\chi(\bg{\alpha})}^2 \nonumber \\
& = I_{LJ}(\rho)
\end{align}
as required.

\section{Decoherence model of Nimmrichter and Hornberger} \label{app:nh_model}
The general form of their modification is rather complex; for simplicity, while highlighting the important features, we take the case where the system may be well approximated by a point-like object of mass $M$ moving in one dimension. Then we can write the modification in a single-particle form
\begin{equation} \label{eqn:NH_model}
\mc{L}_1(\rho) = \frac{1}{\tau} \left( \int\! ds\,dq \; g(s,q) W(s,q) \rho W(s,q)^\dagger - \rho \right),
\end{equation}
where $W(s,q) := e^{i(s\hat{p}-q\hat{x})/\hbar}$ is a displacement of the position and momentum of the object. $\tau$ and $g(s,q)$ both depend on $M$, as well as some chosen reference constants that specify the scale of the problem. The exact form of $g(s,q)$ is not supposed to be essential -- the authors take it to be a normalized Gaussian. Its standard deviations in $s$ and $q$ are $\frac{m_e}{M} \sigma_s$ and $\sigma_q$, where $m_e$, $\sigma_s$ and $\sigma_q$ are some of the reference constants. $m_e$ is chosen to be the electron mass, while the restrictions $\sigma_s \lesssim 20\; \mathrm{pm}$ and $\hbar/\sigma_q \gtrsim 10\; \mathrm{fm}$ are imposed in order to stay within the nonrelativistic regime.

It is simple to show from (\ref{eqn:NH_model}) that when the all the dynamics come from $\mc{L}_1$, then $\partial_t \chi_\rho(r,p) = -\frac{1}{\tau}[1 - \tilde{g}(r,p)] \chi_\rho(r,p)$, where $\tilde{g}(r,p) := \int\! ds\,dq\, g(s,q) e^{-i(ps-rq)/\hbar}$ \cite{nimmrichter2014macroscopic}. Note that the bounded nature of $\tilde{g}(r,p)$ means that the decoherence rate always saturates at some maximum \cite{nimmrichter2013macroscopicity}. Whenever we are far from saturation, we have $\partial_t\chi_\rho \approx -\frac{1}{\tau}\left[ (\frac{m_e \sigma_s r}{M \hbar})^2 + (\frac{\sigma_q p}{\hbar})^2 \right] \chi_\rho(r,p)$. Given that $\frac{1}{\tau} \propto M^2$, it follows that the decoherence model may be well approximated by
\begin{equation}
\partial_t \rho \approx - \text{constant} \times \left[ (m_e \sigma_s)^2 [\hat{x},[\hat{x},\rho]] + (M \sigma_q)^2 [\hat{p},[\hat{p},\rho]] \right].
\end{equation}

\section{Scaling of $I_L$ versus $\mc{F}$} \label{app_scaling}
Here we calculate $\mc{F}(\rho,Z)$, where $Z := \sum_{i=1}^N \sigma^z_i$, for any $N$-qubit state $\rho$ of the form
\begin{align}
\rho & = \sum_{\mb{x},s} p^s_\mb{x} \ketbra{\psi^s_\mb{x}}{\psi^s_\mb{x}}, \\
\ket{\psi^\pm_\mb{x}} & = \frac{1}{\sqrt{2}} \left( \ket{0} \ket{\mb{x}} \pm \ket{1} \ket{\mb{\bar{x}}} \right),
\end{align}
where $\mb{x} = (x_2,x_3,\dots,x_N) \in \{0,1\}^{N-1}$, $\bar{x}:=1-x$ and $s \in \{+,-\}$. Note that the form given above is a spectral decomposition of $\rho$. One can also easily verify the matrix elements
\begin{align}
\braXket{\psi^+_\mb{x}}{Z}{\psi^+_\mb{y}} & = \braXket{\psi^-_\mb{x}}{Z}{\psi^-_\mb{y}} = 0, \nonumber \\
\braXket{\psi^+_\mb{x}}{Z}{\psi^-_\mb{y}} & = (N - 2n(\mb{x}))\, \delta_{\mb{x},\mb{y}},
\end{align}
where $n(\mb{x}) := \sum_{i=2}^N x_i$. Putting this into equation (4) in the main text, we obtain
\begin{equation}
\mc{F}(\rho,Z) = 4\sum_\mb{x} \frac{(p^+_\mb{x} - p^-_\mb{x})^2}{p^+_\mb{x}+p^-_\mb{x}} (N-2n(\mb{x}))^2.
\end{equation}
In particular, when all $p^-_\mb{x} = 0$, we have
\begin{equation}
\mc{F}(\rho,Z) = 4 \sum_\mb{x} p^+_\mb{x} (N-2n(\mb{x}))^2.
\end{equation}
The example state $\rho_N$ in Observation 3 is a special case of the above, so that
\begin{align}
\mc{F}(\rho_N,Z) & = 4 \sum_{k=0}^{N/2-1} \frac{2}{N} (N-2k)^2 \nonumber \\
& = \frac{4}{3}(N+1)(N+2) = O(N^2).
\end{align}
On the other hand, one finds that
\begin{equation}
I_L(\rho_N,Z) = \sum_{k=0}^{N/2} \left( \frac{2}{N} \right)^2 (N-2k)^2 = O(N).
\end{equation}

\end{document}